\documentstyle[12pt]{article}
\input psfig
\def\be{\begin{equation}}
\def\ee{\end{equation}}
\def\ba{\begin{array}}
\def\ea{\end{array}}
\def\beqn{\begin{eqnarray}}
\def\eeqn{\end{eqnarray}}  
\def\bt{\begin{tabular}}
\def\et{\end{tabular}}
\def\bc{\begin{center}}
\def\ec{\end{center}}

\def\vtd{$|V_{td}|$}

\def\as{$a_{\psi}K_S$}
\def\sin2{sin$2\beta$}

\begin{document}
\title{Texture Specific Mass Matrices and CP Violating Asymmetry
in $B^o_d({\bar B}^o_d) \rightarrow \psi K_S$ }
\author{Monika Randhawa and Manmohan Gupta \\
{\it Department of Physics,}\\
{\it Centre of Advanced Study in Physics,}\\
 {\it Panjab University, Chandigarh-
  160 014, India.}}
  \maketitle
\begin{abstract}
  In the context of  texture 4 - zero and texture 5 - zero
  hierarchical quark  mass matrices, CP violating asymmetry
  in $B^o_d({\bar B}^o_d) \rightarrow \psi K_S$ (sin2$\beta$)
  has been evaluated by considering  quark masses at $m_Z$
  scale. For a particular viable texture 4 - zero mass matrix
  the range of \sin2~ is: 0.27 - 0.60 and  for the corresponding
  texture 5 - zero case it is 0.31 - 0.59.  Further our
  calculations reveal a crucial dependence of  sin2$\beta$
  on light quark masses as well as the phase in this sector.
\end{abstract}
The recent  first measurements of time dependent CP asymmetry 
$a_{\psi}K_S$ in $B^o_d({\bar B}^o_d) \rightarrow \psi K_S$ decay
 by BABAR and BELLE collaborations  suggest that
these values could be  smaller than the expectations from Standard Model
analysis of the CKM unitarity triangle. For example, the reported 
asymmetry by BABAR and BELLE are as follows,
\beqn a_{\psi}K_S& =&0.12 \pm 0.37 \pm 0.09~~~~~~~~  {\rm BABAR}~ \cite{babar},
\label{babar} \\
 a_{\psi}K_S&=&0.45^{+0.43~+0.07}_{-0.44~-0.09}~~~~~~~~~~~~~~ {\rm BELLE}~ 
\cite{belle}, \label{belle} \eeqn
whereas the  earlier CDF measurements gave \cite{cdf},
\be a_{\psi}K_S^{{\rm CDF}}=0.79^{+0.41}_{-0.44}, \label{cdf} \ee
and a recent global analysis of the CKM unitarity triangle 
 \cite{parodi} gives the value
\be a_{\psi}K_S^{{\rm SM}} = 0.75 \pm 0.06 . \label{parodi} \ee 
Recently,  several authors
 \cite{kagan} - \cite{buras}   have studied the implications of 
the possibility of low
value of \as~ in comparison to the CDF measurements 
as well as the standard model expectations. 
In particular, Silva and Wolfenstein \cite{silva} have 
examined  the possibilities of physics beyond
the standard model in case \as~$\leq~0.2$. 

In the context of 
texture specific mass matrices,   \as~ has been evaluated
in the leading approximations
\cite{sin2,hall}, however without going into the detailed
implications of \as~ on the texture as well as the mass scale at which 
the quark masses are evaluated.
Recently, it has been demonstrated  \cite{monica}-\cite{frz}
that texture 4 - zero 
quark mass matrices not only  accommodate 
the CKM phenomenology but are also able to reproduce a 
neutrino mixing matrix which can accommodate Solar Neutrino
  Problem, Atmospheric Neutrino Problem and the oscillations observed
  at LSND. In particular, Randhawa {\it et al.} \cite{monica} have shown that
there is a unique set of viable texture 4 - zero mass matrices in the
quark sector as well as in the lepton sector.
The purpose of the present communication
is to investigate in detail and beyond the leading order the 
implications of \as~ measurements for particular viable case of
texture 4 - zero mass matrices as well as for 
texture 5 - zero mass matrices.
 It would also be interesting to examine 
the implications of low values of \as,  in particular of 
\as~ $\leq$ 0.2, a benchmark for Physics beyond the standard model
as advocated by Silva and Wolfenstein \cite{silva}.
 
We begin with the unique set of  texture 4 - zero quark
 mass matrices considered by Randhawa {\it et al.} \cite{monica}, for example,
\vskip 0.2cm
\be 
\bt {cc}
$ M_u=\left(  \ba  {ccc} 0 & A_u & 0\\ A_u^* & D_u & B_u\\
            0 & B_u^* & C_u \ea \right)$ &
$M_d = \left(  \ba  {ccc} 0 & A_d & 0\\ A_d^* & D_d & B_d\\
            0 & B_d^* & C_d \ea \right)$ \\ 
\et  \label{mat}\ee    
\vskip 0.2cm
where $A_u=|A_u|e^{i\alpha_u}$, $A_d=|A_d|e^{i\alpha_d}$,
$B_u=|B_u|e^{i\beta_u}$ and $B_d=|B_d|e^{i\beta_d}.$ The elements
of the mass matrices follow the following mass
 hierarchy \cite{monica,gill,hierarchy}
\be  A_i \ll D_i \sim B_i \ll C_i, ~~~~~~~i=u,d. \label{hier} \ee

Within the Standard Model, \as~ is related to the angle $\beta$ of the
unitarity triangle,  expressed as
\be a_{\psi}K_S^{{\rm SM}} = {\rm sin}2\beta,~~~~\beta \equiv {\rm arg}
\left[ \frac{V_{cd}V_{cb}^{*}}{V_{td}V_{tb}^{*}}\right], \label{beta} \ee
 \sin2~ can be calculated by evaluating the elements
$V_{cd}$, $V_{cb}$, $V_{td}$ and   $V_{tb}$
from the above mass matrices.

The above matrices can be diagonalized
exactly and  the corresponding  CKM matrix elements
 can  easily be found, for details we refer
the reader to reference \cite{gill}. However for the sake
of readability of manuscript as well as for facilitating the 
discussion to evaluate \sin2, we reproduce below the exact expressions
of $V_{cd}$, $V_{cb}$, $V_{td}$ and $V_{tb}$.
\beqn 
 V_{cd}&=&-ae^{-i\phi_1} + c{\sqrt{(1-R_u)(1-R_d)}} + 
c{\sqrt{(b^2+R_u)(d^2+R_d)}}e^{i\phi_2}, \label{vcd} \\
 V_{cb}&=& -acd^2{\sqrt\frac{d^2+R_d}{1-R_d}}e^{-i\phi_1}
    + {\sqrt{(1-R_u)(d^2+R_d)}} \nonumber \\
 & & - {\sqrt{(b^2+R_u)(1-R_d)}}e^{i\phi_2}, \label{vcb} \\
 V_{td}&=&ab ^2{\sqrt\frac{b^2+R_u}{1-R_u}}e^{-i\phi_1} +
   c{\sqrt{(b^2+R_u)(1-R_d)}} \nonumber \\
 & &  - c{\sqrt{(1-R_u)(d^2+R_d)}}e^{i\phi_2}, \label{vtd} \\ 
 V_{tb}&=&acb^2d^2{\sqrt\frac{(b^2+R_u)(d^2+R_d)}{(1-R_u)(1-R_d)}}
     e^{-i\phi_1} + {\sqrt{(b^2+R_u)(d^2+R_d)}} \nonumber \\
 & & + {\sqrt{(1-R_u)(1-R_d)}}e^{i\phi_2},
    \label{vtb} 
\eeqn 
where  $a=\sqrt{m_u/m_c}$, $b=\sqrt{m_c/m_t}$,
 $c=\sqrt{m_d/m_s}$, $d=\sqrt{m_s/m_b}$,
 $\phi_1=\alpha_u-\alpha_d.$, $\phi_2=\beta_u-\beta_d.$
 $R_u=D_u/m_t$ and  $R_d=D_d/m_b$.
~In principle, \sin2~ can be calculated using equations 
 \ref{beta}-\ref{vtb}, however before doing that we first 
ensure that by varying the various  input parameters, the
CKM matrix elements are within their respective range
given by PDG  \cite{pdg}. In carrying out these calculations,
we have taken  the quark masses at $m_Z$ scale as recently
advocated by Fusaoka and Koide \cite{koide} as well as by Fritzsch
 and Xing \cite{frz}. For the sake of completion, however, we have
 also repeated the whole analysis with masses at 1 GeV scale, the scale 
 conventionally used.

To facilitate the analysis, without loss
of generality we first consider $\phi_2=0$ as advocated
by several authors \cite{monica,frz}.
As mentioned earlier we have carried out our calculations at 
two different mass scales i.e. at $m_Z$ scale and at 1 GeV,
 the corresponding input masses are 
summarized in Table \ref{tabinp}. For calculating the limits on \sin2,
we scanned the full ranges of all the input masses at different CLs 
as well as at both the
scales, varying $\phi_1$ from $0^o$ to $180^o$. It may be of interest
to point out that while carrying out the variations in 
$R_u$ and $R_d$, we have restricted their variation upto 0.2 
only as the values higher than that are not able to
 reproduce the CKM elements within
their range given by PDG. This is in accordance with our earlier
calculations \cite{monica} as well as the hierarchical
 structure of mass matrices 
 described by equation \ref{hier}. 
Having taken care of the CKM matrix elements being within the limits
mentioned by PDG, we proceed to find a range for \sin2~ 
using expressions \ref{beta} - \ref{vtd}.
A similar exercise is carried out for 
(i) $D_u$ = 0, $D_d~ {\not=}~0 $, (ii) $D_u~ {\not=}~0$, $D_d$
 = 0, the two cases corresponding to texture 5 - zero matrices.

\begin{figure}
   \centerline{\psfig{figure=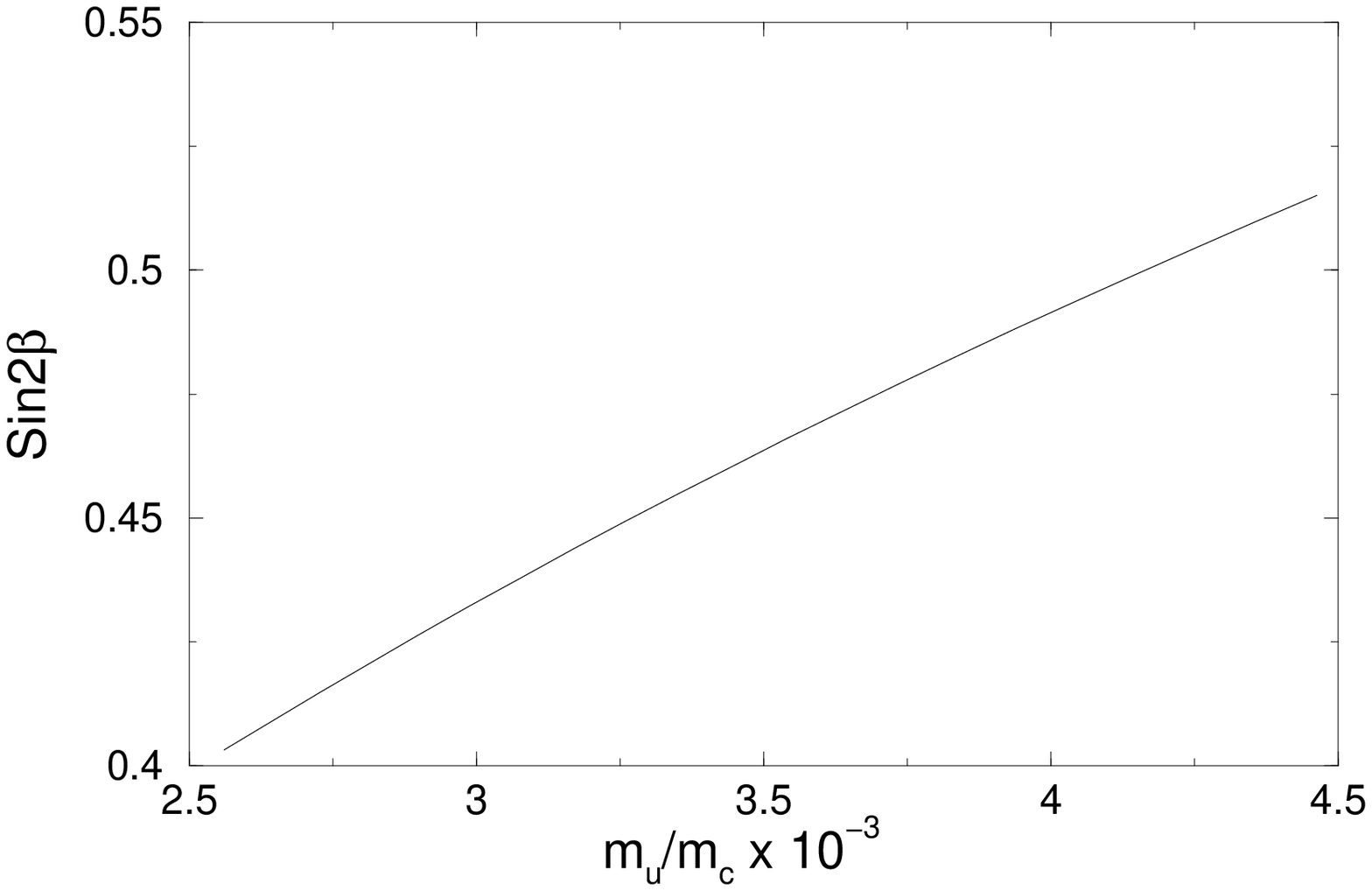,width=3in,height=3in}}
   \caption{Variation of sin2$\beta$ with $m_u/m_c$ at $m_Z$ scale.
 All other  masses are at their mean values,
 whereas $\phi_1=90^o$,   $\phi_2=0$ and $R_u=R_d=0.1$.}
  \label{figa}
  \end{figure}

\begin{figure}
   \centerline{\psfig{figure=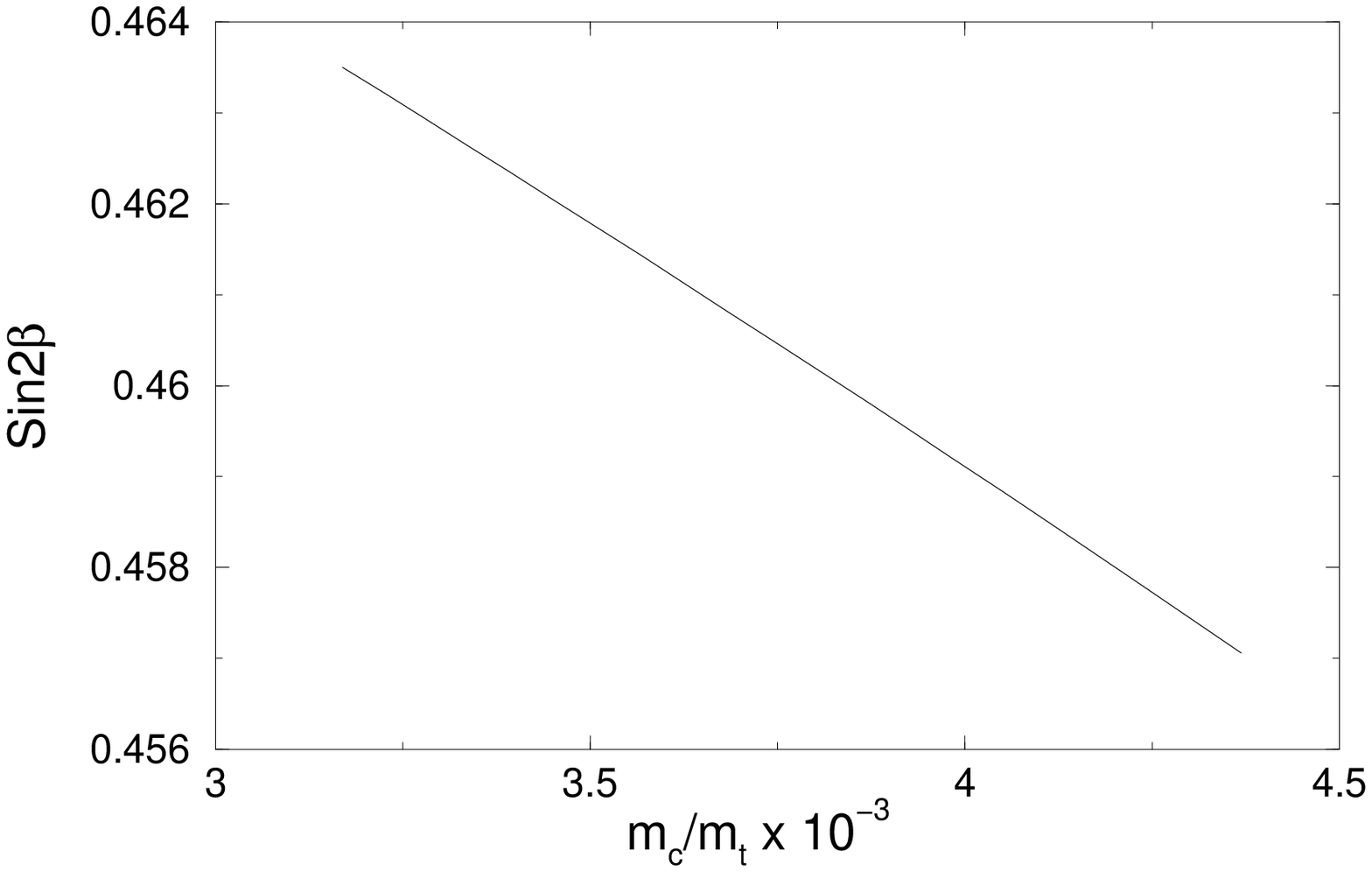,width=3in,height=3in}}
   \caption{Variation of sin2$\beta$ with $m_c/m_t$ at $m_Z$ scale.
 All other  masses are at their mean values, whereas $\phi_1=90^o$, 
  $\phi_2=0$ and $R_u=R_d=0.1$.}
  \label{figb}
  \end{figure}

\begin{figure}
   \centerline{\psfig{figure=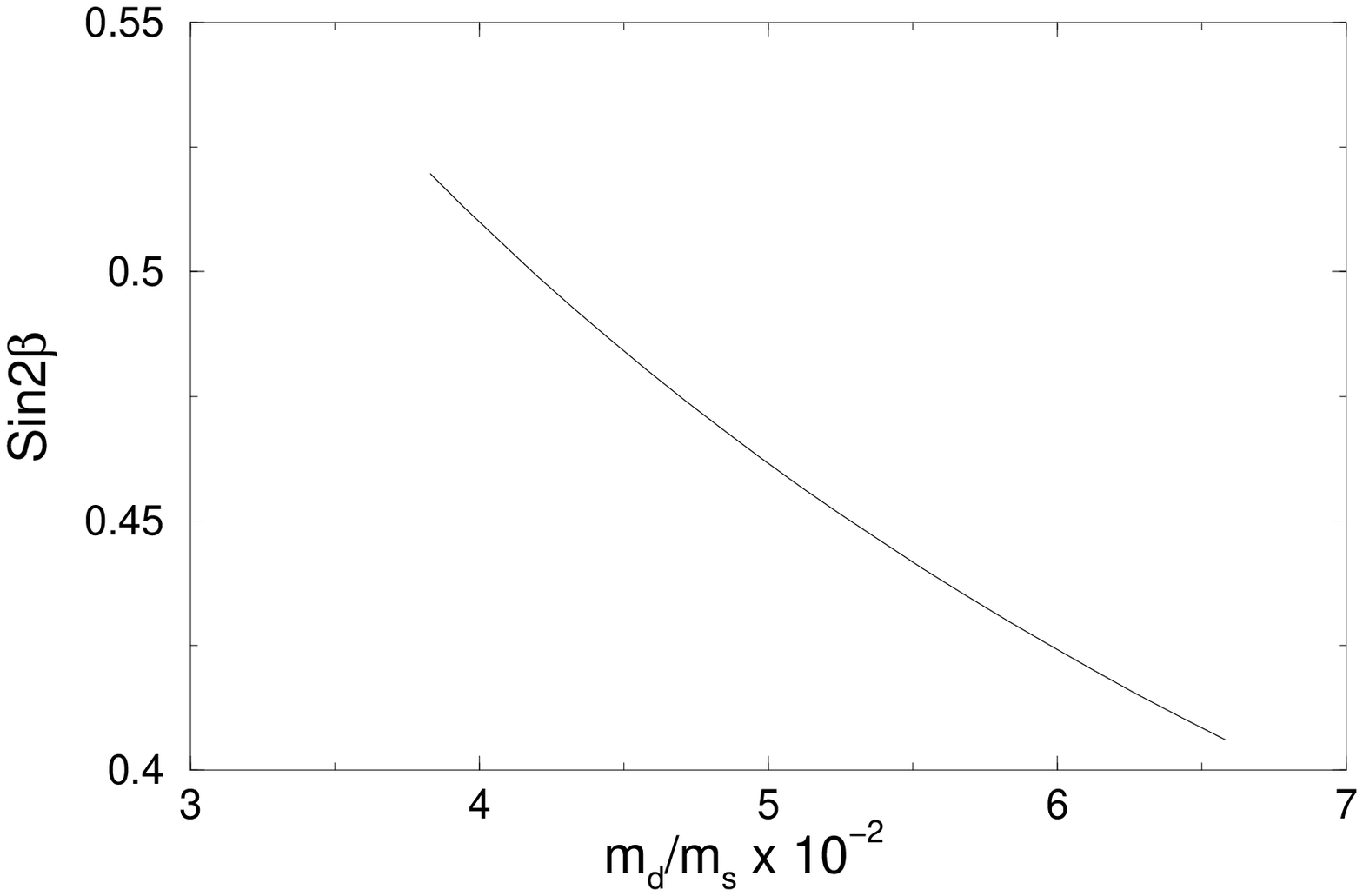,width=3in,height=3in}}
    \caption{Variation of sin2$\beta$ with $m_d/m_s$
  at $m_Z$ scale. All other masses are at their mean values, 
whereas $\phi_1=90^o$,    $\phi_2=0$ and $R_u=R_d=0.1.$}
\label{figc}
  \end{figure}

\begin{figure}
   \centerline{\psfig{figure=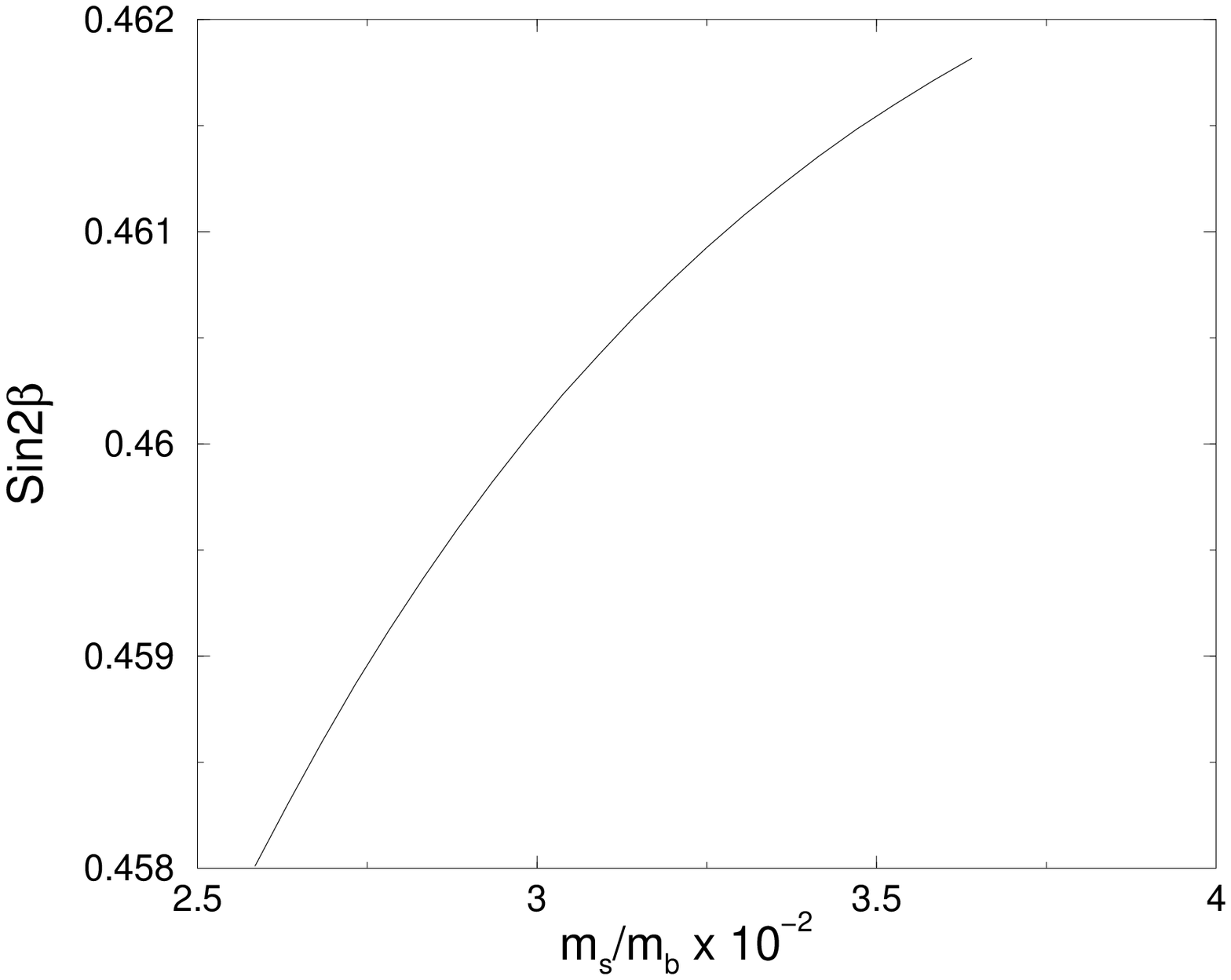,width=3in,height=3in}}
  \caption{Variation of sin2$\beta$ with $m_s/m_b$
  at $m_Z$ scale.  All other
  masses are at their mean values, whereas $\phi_1=90^o$, 
  $\phi_2=0$ and $R_u=R_d=0.1$.}
  \label{figd}
  \end{figure}

In Table \ref{tab1}, we have summarized the  results
of our calculations at the different mass scales for 
  texture 4 - zero and texture 5 - zero mass  matrices.
From the table one can immediately find that the range of \sin2~
in the case of texture 4 - zero mass matrices, with
input masses at $m_Z$ scale and  at 1$\sigma$ CL, is given by 
\be sin2\beta =  0.27 -  0.60 . \label{sinmzb1} \ee
This range looks to be narrow in comparison with the 
BELLE and BABAR results and is ruled out by the SM analysis.
 The corresponding range for \sin2~ narrows further when quark masses 
are considered at  1 GeV scale, for example,
\be sin2\beta = 0.39  -  0.54. \label{sinb1} \ee
This can be easily understood from the fact that the light quark masses 
at  $m_Z$ scale show much more scatter compared to masses at 1 GeV scale.
 In view of the sensitive dependence of \sin2~ on the quark masses,
  in figures \ref{figa} - \ref{figd},
 we have plotted the variation of \sin2~
with mass ratios $m_u/m_c$, $m_d/m_s$, 
$m_c/m_t$ and $m_s/m_b$.
From these figures, it is easy to conclude that 
 \sin2~ is very sensitively dependent 
on the ratios of the light quark masses,
$m_u/m_c$ and $m_d/m_s$, while variations in
$m_c/m_t$ and $m_s/m_b$ do not affect \sin2~ much.
This gets further emphasized when one closely examines
the figures, for example, \sin2~
varies from 0.40 to 0.52 when $m_u/m_c$ varies from 0.0026 to 0.0045, 
while it varies only from 0.464 to 0.457  when $m_c/m_t$ varies from
0.0032 to 0.0044. Similarly  \sin2~
varies from 0.52 to 0.41 when $m_d/m_s$ varies from 0.0383  to 0.0658,  
while it varies only from 0.458 to 0.461 when $m_s/m_b$ varies from
0.0258 to 0.0364. It is perhaps desirable to mention that 
while considering the above mentioned variation of \sin2~ on a
 given mass ratio all other masses have been kept at their mean
 values at $m_Z$ scale, whereas
 $\phi_1 = 90^o$ and $R_u=R_d=0.1$. 

 In view of the scale sensitivity of
\sin2, it is perhaps desirable to study the affect of quark masses 
on \sin2~ at higher confidence levels of quark masses in comparison to the
$1 \sigma$ CL corresponding to equations   \ref{sinmzb1} - \ref{sinb2}.
In the table \ref{tab1}, we have also listed the results for \sin2~ 
with the input quark masses being at their 2$\sigma$ and 3$\sigma$
 confidence levels. A look at the table reveals that
when the quark masses are considered  at $2 \sigma$ CL,
we obtain the following ranges for \sin2~ for the set of texture 4 - zero 
matrices given in equation \ref{mat}:  0.057 - 0.68.
 These ranges get further broadened when masses are considered
 at their $3 \sigma$ CL, for example, 0.04 - 0.75.
Thus we see that with input masses at their $2 \sigma$
and  $3 \sigma$ CL, the entire range of
BABAR and BELLE is covered,
once again emphasizing the sensitivity of \sin2~ on the quark masses.
This brings into focus the better evaluation of 
light quark masses.

Further scrutiny of the Table \ref{tab1} reveals interesting
results for texture 5 - zero case.
 For example, we obtain the following range for
sin2$\beta$ with   $D_u~ {\not=}~0$, $D_d$  = 0 
and with quark masses at  $m_Z$ scale and at 1$\sigma$ CL,
\be sin2\beta = 0.31 - 0.59.  \label{sinmzb2} \ee
and correspondingly with the masses at 1 GeV we get,
\be sin2\beta = 0.45 -  0.54  \label{sinb2} \ee
Thus,  in comparison to the corresponding ranges for texture 4 - zero
 matrices, the lower bound on \sin2~ goes up
somewhat while there is not much change in the upper bound.
This can be understood by an examination of equations
 \ref{beta} and  \ref{vtd} where $D_d=0$ results in
lowering  the upper bound on \vtd,
thus pushing up the lower bound on \sin2.
In the other texture 5 - zero case, for example $D_d~ {\not=}~0$
and $D_u$  = 0, we find that it is not meaningful to talk of the
range of \sin2 as in this case the CKM matrix elements do not show
 overlap with the PDG CKM matrix
even after the full variation of all the parameters.

  A few comments are in order. In view of the dependence of
  \sin2~ on $\phi_1$ and $\phi_2$
  through $V_{cd}$, $V_{cb}$, $V_{td}$ and   $V_{tb}$,
  we have also studied the case when both $\phi_1$ and $\phi_2$ are
  taken non zero. The results in this case do not
  show much deviation from the case when $\phi_2 = 0$ and
  $\phi_1$ is varied. However when  $\phi_1 = 0$ and
  $\phi_2$  is given full variation, interestingly
  we find that we are not able to reproduce the CKM matrix elements
  and hence finding a range for \sin2~ in this case is meaningless.
  Thus, it seems that the CP violating phase resides only in the
light quark sector,  in agreement with the conclusions of Fritzsch and Xing
  \cite{sin2} based on leading order calculations only.

Interestingly, from equations \ref{sinmzb1} - \ref{sinb2}, we find that 
  a value of \sin2~ lower than 0.2 would rule out, 
with good deal of confidence, texture 4 - zero
  and texture 5 - zero matrices. This would force 
   one to consider texture 3 - zero and texture 2 - zero mass matrices,
which would be discussed elsewhere.
Therefore it seems that a sharper measurement of \sin2~ will have strong
  bearings on the specific textures of mass matrices.

  To conclude, we have found a range for
\sin2~ using texture 4 - zero and texture 5 - zero hierarchical quark 
mass matrices with input quark masses at $m_Z$ scale.
In the texture 4 - zero case with masses at 1$\sigma$ CL, we get
\sin2=0.27 - 0.60 and in the texture 5 - zero case we get \sin2
=0.31 - 0.59. Both texture 5 - zero and texture 4 - zero matrices 
are ruled out if \sin2~ is found to be $\leq$ 0.2 and one may have
to go to texture 3 - zero matrices. Our analysis
indicates a sensitive dependence  of \sin2~
on the light quark masses as well as the phase in this sector.
\vskip 1cm
  {\bf ACKNOWLEDGMENTS}\\
M.G. would like to thank S.D. Sharma for useful discussions.
M.R. would like to thank CSIR, Govt. of India, for
 financial support and also the Chairman, Department of Physics,
for providing facilities to work in the department.

\begin{table}[h]
\begin{center}
\bt{|c|c|c|} \hline
& At $\mu$=1~GeV & At $\mu = m_Z$ \\ \hline
& & \\
$m_u$ & 0.00488 $\pm$ 0.00057 & $0.00233^{+0.00042}_{-0.00045}$ \\
& & \\
$m_d$ & 0.00981 $\pm$ 0.00065 & $0.00469^{+0.00060}_{-0.00066}$ \\
& & \\
$m_s$ & 0.1954 $\pm$ 0.0125 & $0.0934^{+0.0118}_{-0.0130}$ \\
& & \\
$m_c$ & $1.506^{+0.048}_{-0.037}$  & $0.677^{+0.056}_{-0.061}$ \\
& & \\
$m_b$ & $7.18^{+0.59}_{-0.44}$ & 3.0 $\pm$ 0.11 \\
& & \\
$m_t$ & $475^{+86}_{-71}$ & 181 $\pm$ 13 \\ 
& & \\ \hline \et
\caption{Running quark masses $m_q(\mu)$ 
(in units of GeV)}\cite{koide}. 
\label{tabinp}
\end{center}
\end{table}
\begin{table}
\bt{|c|c|c|c|c|} \hline \hline
& \multicolumn{2}{c|} {masses at $\mu=m_Z$} &
   \multicolumn{2}{c|}{masses at $\mu$=1 GeV} \\ \cline{2-5}
 & texture 4 zeros &  \bt{c} texture 5 zeros \\ ($D_d=0$,  $D_u{\not =0}$)
\\ \et & texture 4 zeros & \bt{c} texture 5 zeros\\  ($D_d=0$,  
$D_u{\not =0}$) \\ \et \\ \hline
 \bt{c}  Sin2$\beta$ \\ (with quark
 masses\\ at 1$\sigma$ CL) \\ \et & 0.27 - 0.60 &  0.31 - 0.59 
 & 0.39 - 0.54 & 0.45 - 0.54  \\ \hline
 \bt{c} Sin2$\beta$ \\ (with quark
 masses\\ at 2$\sigma$ CL) \\ \et & 0.057 - 0.68  &  0.08 - 0.65 
 & 0.27 - 0.57 & 0.38 - 0.56 \\ \hline
  \bt{c} Sin2$\beta$ \\ (with quark
 masses \\ at 3$\sigma$ CL) \\ \et  & 0.04 - 0.75 &  0.05 - 0.73 
 & 0.06 - 0.61 & 0.07 - 0.58 \\ \hline
\et
\caption{The range for \sin2~at different confidence levels of quark masses.}
\label{tab1}
\end{table}
\pagebreak
\end{document}